\documentclass[aps,prd,amsmath, amssymb,eqsecnum,
nofootinbib,showpacs,tightenlines]{revtex4}

\usepackage{bm}
\usepackage{verbatim} 

\newcommand{\be}{\begin{equation}}
\newcommand{\ee}{\end{equation}}
\newcommand{\bea}{\begin{eqnarray}}
\newcommand{\eea}{\end{eqnarray}}
\newcommand{\nn}{\nonumber}
\begin{document}

\title{How does Casimir energy fall? II. Gravitational acceleration
of quantum vacuum energy}
\author{Kimball A. Milton} 
\email{milton@nhn.ou.edu}
\homepage{http://www.nhn.ou.edu/
\author{Prachi Parashar}
\email{prachi@nhn.ou.edu}
\author{K. V. Shajesh}
\email{shajesh@nhn.ou.edu}
\homepage{http://www.nhn.ou.edu/
\author{Jef Wagner}
\email{wagner@nhn.ou.edu}
\affiliation{Oklahoma Center for High Energy Physics 
and Homer L. Dodge Department of Physics and Astronomy,
University of Oklahoma, Norman, OK 73019, USA}

\date{\today}


\begin{abstract}
It has been demonstrated that quantum vacuum energy gravitates according
to the equivalence principle, at least for the finite Casimir energies
associated with perfectly conducting parallel plates.  We here add further
support to this conclusion by considering parallel semitransparent plates,
that is, $\delta$-function potentials, acting on a massless scalar field,
in a spacetime defined by Rindler coordinates $(\tau,x,y,\xi)$.  
Fixed $\xi$ in such a spacetime represents
uniform acceleration.  We calculate the force on systems consisting of one or
two such plates at fixed values of $\xi$.  
In the limit of large Rindler coordinate $\xi$ (small 
acceleration), we recover (via the equivalence principle)
the situation of weak gravity, and find that the gravitational force on
the system is just $M\mathbf{g}$, where $\mathbf{g}$ 
is the gravitational acceleration
and $M$ is the total mass of the system, consisting of the mass of the plates
renormalized by the Casimir energy of each plate separately, plus the energy
of the Casimir interaction between the plates.  This reproduces the previous
result in the limit as the coupling to the $\delta$-function potential
approaches infinity.
\end{abstract}

\maketitle

\section{Introduction}
The subject of quantum vacuum energy, or of Casimir energy, has engendered
a certain controversy from the beginning because of the presence of
divergences, which make it difficult to extract self-energies for single
bodies \cite{Deutsch:1978sc,Candelas:1981qw,Graham:2003ib,barton}.
Although it appears that many of these divergences can be consistently
isolated when calculating the Casimir forces between distinct bodies,
the issue of how divergent and finite Casimir energies couple to gravity
remains unclear.  

In the last few years there have been several calculations of the gravitational
acceleration imparted to the Casimir energy associated 
with a pair of perfectly
conducting plates \cite{karim,calloni,caldwell,Sorge:2005ed,bimonte}.  The
results were inconsistent, and there was no consensus that the 
gravitational force agreed with the equivalence
principle.  Recently, we have shown that indeed the gravitational force
on a Casimir apparatus 
is exactly that required by the equivalence principle, that is,
that the gravitational mass of the Casimir energy is just the Casimir energy
itself \cite{Fulling:2007xa}.  Other authors now  agree with our
conclusion \cite{Bimonte:2007zt}.  However, that calculation included only
the finite Casimir energy of the two plates, so the question of what happens
to the divergent contributions remains unanswered.

Here we answer that question.  We describe a uniformly accelerated system by
Rindler coordinates \cite{rindler1966}, which naturally represent frames
undergoing hyperbolic motion.  We consider both a single plate, and two
plates, both represented by $\delta$-function potentials, what are sometimes
called semitransparent plates.  In Minkowski space, the Casimir energies
for such systems have been considered by many authors
\cite{Bordag:1992cm,Graham:2003ib,Milton:2004vy,Milton:2004ya,khus}.
Saharian et al.~\cite{Saharian:2003fd} considered Dirichlet, Neumann, and 
perfectly conducting plates in Rindler coordinates, and showed for
rigid acceleration of those plates, in the limit
of large Rindler coordinate, which corresponds to the weak gravitational
field limit,  that the finite Casimir energy undergoes the normal
acceleration.  We carry out that calculation here for
semitransparent plates (which reduce to Dirichlet plates in the
strong-coupling limit) and find that
both for a single plate, and for two parallel plates (both orthogonal
to the Rindler spatial coordinate) both the divergent and finite parts
of the Casimir energy gravitate according to the equivalence principle,
and that the divergent energies serve to renormalize the 
inertial and gravitational masses of each separate plate.\footnote{Acceleration
of semitransparent plates has been considered by many authors in connection
with quantum radiation 
\cite{Barton:1995iw,Calogeracos:1995ix,Helfer:2000fg,Nicolaevici:1999ga,
Haro:2007ue}.  Gravitational effects of surface energies in Rindler and
de Sitter spacetimes are considered in 
Refs.~\cite{saharian04,setare05,saharian06}.}

\section{Green's Functions in Rindler Coordinates} 
\label{cg}
Relativistically, uniform acceleration is described by hyperbolic
motion,
\be
z = \xi \cosh \tau\quad\text{and} \quad t = \xi \sinh \tau.
\label{hy-eq}
\ee
Here the proper acceleration of the particle described by these
equations is $\xi^{-1}$, and we have
chosen coordinates so that at time $t=0$, $z(0)=\xi$.
Here we are going to consider the corresponding metric
\begin{equation}
ds^2 = - dt^2 + dz^2 + dx^2 + dy^2 
     = - \xi^2 d\tau^2 + d\xi^2 + dx^2 + dy^2.
\end{equation}
In these coordinates, the d'Alembertian operator takes on cylindrical form
\be
-\left(\frac\partial{\partial t}\right)^2
+\left(\frac\partial{\partial z}\right)^2+\bm{\nabla}_\perp^2
=-\frac1{\xi^2}\left(\frac\partial{\partial\tau}\right)^2+\frac1\xi
\frac\partial{\partial\xi}\left(\xi\frac\partial{\partial\xi}\right)
+\bm{\nabla}_\perp^2,
\ee
where $\perp$ refers to the $x$-$y$ plane.

\subsection{Green's function for one plate}
For a scalar field in these coordinates, subject to a potential $V(x)$,
the action is
\begin{equation}
W = \int d^4x \sqrt{-g(x)}
{\cal{L}}(\phi(x)),\label{action}
\end{equation}
where $x\equiv(\tau,x,y,\xi)$ represents the coordinates, 
$d^4x = d\tau \,d\xi\, dx\, dy$ is the coordinate volume element,
$g_{\mu\nu}(x) = \mbox{diag}(-\xi^2,+1,+1,+1)$ defines the metric,
$g(x) = \det g_{\mu\nu}(x)=-\xi^2$ is the determinant of the metric, 
and the Lagrangian density is
\begin{equation}
{\cal{L}}(\phi(x)) 
= - \frac{1}{2} g_{\mu\nu}(x) \partial^\mu \phi(x)\partial^\nu \phi(x)
  - \frac{1}{2} V(x) \phi(x)^2,
\end{equation}
where for a single semitransparent plate located at $\xi_1$
\begin{equation}
V(x) = \lambda \delta(\xi - \xi_1),
\end{equation}
and $\lambda>0$ is the coupling constant having dimensions of mass.
More explicitly we have
\begin{equation}
W = \int d^4x \,\frac{\xi}{2}
\left[\frac{1}{\xi^2} 
 \left( \frac{\partial \phi}{\partial \tau} \right)^2
- \left( \frac{\partial \phi}{\partial \xi} \right)^2
- \left( {\bm{\nabla}}_\perp \phi \right)^2
- V(x) \phi^2
\right].
\end{equation}
Stationarity of the action under an arbitrary variation in 
the field leads to the equation of motion 
\begin{equation}
\left[
- \frac{1}{\xi^2} \frac{\partial^2}{\partial \tau^2}
+ \frac{1}{\xi} \frac{\partial}{\partial \xi} 
  \xi \frac{\partial}{\partial \xi}
+ {\bm{\nabla}}_\perp^2 - V(x)
\right] \phi(x) = 0.\label{eom}
\end{equation}

The corresponding Green's function satisfies the differential equation
\begin{equation}
-\left[
- \frac{1}{\xi^2} \frac{\partial^2}{\partial \tau^2}
+ \frac{1}{\xi} \frac{\partial}{\partial \xi}
  \xi \frac{\partial}{\partial \xi}
+ {\bm{\nabla}}_\perp^2 - V(x)
\right] G(x,x^\prime) =
\frac{\delta(\xi-\xi^\prime)}{\xi}
\delta(\tau - \tau^\prime) \delta({\bf x}_\perp-{\bf x}_\perp^\prime).
\end{equation}
Since in our case $V(x)$ has only $\xi$ dependence we 
can write this in terms of the reduced Green's function $g(\xi, \xi')$,
\begin{equation}
G(x,x^\prime)
= \int_{-\infty}^{\infty}
  \frac{d\omega}{2 \pi} \int \frac{d^2 \mathbf{k}_\perp}{(2 \pi)^2}
  e^{-i \omega (\tau - \tau^\prime)}
  e^{i {\bf k}_\perp \cdot ({\bf x} - {\bf x}^\prime)_\perp}
  g(\xi,\xi^\prime),
\label{capG}
\end{equation}
where $g(\xi,\xi^\prime)$ satisfies
\begin{equation}
-\left[
\frac{1}{\xi} \frac{\partial}{\partial \xi}
  \xi \frac{\partial}{\partial \xi}
+\frac{\omega^2}{\xi^2}
- k_\perp^2 - V(x)
\right] g(\xi,\xi^\prime) = \frac{\delta(\xi-\xi^\prime)}{\xi}.
\label{zgf}
\end{equation}

We recognize this equation as defining the semitransparent cylinder
problem \cite{Cavero-Pelaez:2006rt}, with the replacements
\be
m\to\zeta=-i\omega,\quad \kappa\to k=k_\perp,
\ee
so that we may immediately write down the solution in terms of modified
Bessel functions,
\begin{subequations}
\bea
g(\xi,\xi')&=&I_\zeta(k\xi_<)K_\zeta(k\xi_>)
-\frac{\lambda\xi_1 K_\zeta^2(k\xi_1)
I_\zeta(k\xi)I_\zeta(k\xi')}{1+\lambda\xi_1 I_\zeta(k\xi_1)K_\zeta(k\xi_1)},
\qquad \xi,\xi'<\xi_1,\\
&=&I_\zeta(k\xi_<)K_\zeta(k\xi_>)-\frac{\lambda \xi_1 I_\zeta^2(k\xi_1)
K_\zeta(k\xi)K_\zeta(k\xi')}{1+\lambda\xi_1 I_\zeta(k\xi_1)K_\zeta(k\xi_1)},
\qquad \xi,\xi'>\xi_1.
\eea
\end{subequations}
Note that in the strong-coupling limit, $\lambda\to\infty$, this reduces
to the Green's function satisfying Dirichlet boundary conditions at 
$\xi=\xi_1$.

\subsection{Minkowski-space limit}

To recover the Minkowski-space Green's function for the semitransparent plate,
we use the uniform asymptotic expansion (Debye expansion), based on the
limit
\bea
\xi\to\infty,&&\quad \xi_1\to\infty, \quad \xi-\xi_1 \mbox{ finite },
\quad \zeta\to\infty,\quad \zeta/\xi_1 \mbox{ finite }.
\eea
For large $\zeta$
\be
I_\zeta(\zeta z)
\sim \sqrt{\frac{t}{2\pi\zeta}} \, e^{\zeta \eta(z)}
\sum_{n=0}^{\infty} \frac{1}{\zeta^n} u_n(t),
\quad
K_\zeta(\zeta z)
\sim \sqrt{\frac{\pi t}{2\zeta}} \, e^{-\zeta \eta(z)}
\sum_{n=0}^{\infty} \frac{(-1)^n}{\zeta^n} u_n(t),
\label{uae}
\ee
where
\be
t=\frac{1}{\sqrt{1 + z^2}}
\quad\text{and} \quad
\eta(z) = \sqrt{1 + z^2}
+ \ln \left[\frac{z}{1 + \sqrt{1+z^2}}
\right], 
\ee
and $u_n(t)$ are polynomials of order $3n$ in $t$ \cite{maias}.
Here $z\zeta=k\xi$, for example.
Expanding the above expressions around some arbitrary point $\xi_0$, 
chosen such that the differences 
$\xi - \xi_0$, $\xi'-\xi_0$, and $\xi_1 - \xi_0$ are finite, we
find for the leading term, for example,
\begin{equation}
\sqrt{\xi \xi'} \, I_\zeta(k\xi) K_\zeta(k\xi')
\sim \frac{1}{2 \kappa} \, e^{\kappa (\xi - \xi')},
\label{IKap}
\end{equation}
where $\kappa^2 = k^2 + \hat\zeta^2$,  $\hat\zeta=\zeta/\xi_0$.
In this way, taking for simplicity $\xi_0=\xi_1$, we find the Green's function
for a single plate in Minkowski space,
\begin{equation}
\xi_1g(\xi,\xi')\to g^{(0)}(\xi,\xi^\prime) =
\frac{1}{2 \kappa} \,e^{-\kappa |\xi - \xi^\prime|}
- \frac{\lambda}{\lambda + 2 \kappa}
  \frac{1}{2 \kappa} 
  \, e^{-\kappa|\xi - \xi_1|} e^{-\kappa|\xi^\prime - \xi_1|}.
\label{1-g0-zzp}
\end{equation}

\subsection{Green's function for two parallel plates}\label{sec2c}
For two semitransparent plates perpendicular to the $\xi$-axis
and located at $\xi_1$, $\xi_2$, with couplings $\lambda _1$ and
$\lambda_2$, respectively, we find the following form for
the Green's function:
\begin{subequations}
\bea
g(\xi,\xi')&=&I_<K_>
-\frac{\lambda_1\xi_1K_1^2+\lambda_2\xi_2K_2^2-\lambda_1\lambda_2\xi_1\xi_2
K_1K_2(K_2I_1-K_1I_2)}{\Delta}II_\prime,\quad \xi, \xi'<\xi_1,\\
&=&I_<K_>-\frac{\lambda_1\xi_1I_1^2
+\lambda_2\xi_2I_2^2+\lambda_1\lambda_2\xi_1\xi_2
I_1I_2(I_2K_1-I_1K_2)}{\Delta}KK_\prime,\quad \xi, \xi'>\xi_2,\\
&=&I_<K_>
-\frac{\lambda_2\xi_2K_2^2(1+\lambda_1\xi_1K_1I_1)}\Delta II_\prime\nn\\
&&\quad\mbox{}-\frac{\lambda_1\xi_1I_1^2(1+\lambda_2\xi_2K_2I_2)}
\Delta KK_\prime
+\frac{\lambda_1\lambda_2\xi_1\xi_2I_1^2K_2^2}\Delta(IK_\prime+KI_\prime),
\quad \xi_1<\xi,\xi'<\xi_2,
\eea
\end{subequations}
where
\be
\Delta
=(1+\lambda_1\xi_1K_1I_1)(1+\lambda_2\xi_2K_2I_2)-\lambda_1\lambda_2\xi_1\xi_2
I_1^2K_2^2,\label{Delta}
\ee
and we have used the abbreviations
$I_1=I_\zeta(k\xi_1)$, $I=I_\zeta(k\xi)$, $I_\prime=I_\zeta(k\xi')$, etc.

Again we can check that these formulas reduce to the well-known 
Minkowski-space limits.  In the $\xi_0\to\infty$ limit, 
the uniform asymptotic expansion (\ref{uae}) gives, for $\xi_1<\xi,\xi'<\xi_2$
\bea
\xi_0g(\xi,\xi')\to g^{(0)}(\xi,\xi')&=&\frac1{2\kappa}e^{-\kappa|\xi-\xi'|}
+\frac1{2\kappa\tilde\Delta}\bigg[\frac{\lambda_1\lambda_2}
{4\kappa^2}2\cosh\kappa(\xi-\xi')\nn\\
&&\mbox{}-\frac{\lambda_1}{2\kappa}\left(1+\frac{\lambda_2}{2\kappa}\right)
e^{-\kappa(\xi+\xi'-2\xi_2)}
-\frac{\lambda_2}{2\kappa}\left(1+\frac{\lambda_1}{2\kappa}\right)
e^{\kappa(\xi+\xi'-2\xi_1)}\bigg],
\eea
where ($a=\xi_2-\xi_1$)
\be
\tilde\Delta=\left(1+\frac{\lambda_1}{2\kappa}\right)
\left(1+\frac{\lambda_2}{2\kappa}\right)e^{2\kappa a}
-\frac{\lambda_1\lambda_2}{4\kappa^2},\label{tDelta}
\ee
which is exactly the expected result \cite{Milton:2004vy}.  
The correct limit is also obtained in the other two regions.

\section{Gravitational Acceleration of Casimir Apparatus}

We next consider the situation when the plates are forced to 
``move rigidly'' \cite{born} in such a way that the proper distance 
between the plates is preserved. This is achieved if the 
two plates move with different but 
constant proper accelerations.

The canonical energy-momentum or stress tensor derived from the
action (\ref{action}) is
\begin{equation}
T_{\alpha \beta}(x)
= \partial_\alpha \phi(x) \partial_\beta \phi(x)
+ g_{\alpha \beta}(x) {\cal{L}}(\phi(x)),
\end{equation}
where the Lagrange density includes the $\delta$-function potential. The  
components referring to the pressure and the energy density are
\begin{subequations}
\begin{eqnarray}
T_{33}(x)
&=&  \frac{1}{2}\frac{1}{\xi^2}
 \left( \frac{\partial \phi}{\partial \tau} \right)^2
+ \frac{1}{2} \left( \frac{\partial \phi}{\partial \xi} \right)^2
- \frac{1}{2} \left( {\bm{\nabla}}_\perp \phi \right)^2
- \frac{1}{2} V(x) \phi^2
\\
\frac{1}{\xi^2} \, T_{00}(x)
&=&  \frac{1}{2}\frac{1}{\xi^2}
 \left( \frac{\partial \phi}{\partial \tau} \right)^2
+ \frac{1}{2} \left( \frac{\partial \phi}{\partial \xi} \right)^2
+ \frac{1}{2} \left( {\bm{\nabla}}_\perp \phi \right)^2
+ \frac{1}{2} V(x) \phi^2.
\end{eqnarray}
\end{subequations}
The latter may be written in an alternative convenient form using
the equations of motion (\ref{eom}):
\be
T_{00}=\frac12\left(\frac{\partial\phi}{\partial\tau}\right)^2-\frac12\phi
\frac{\partial^2}{\partial\tau^2}\phi+\frac\xi 2\frac\partial{\partial\xi}
\left(\phi\xi\frac\partial{\partial\xi}\phi\right)
+\frac{\xi^2}2\bm{\nabla}_\perp\cdot(\phi\bm{\nabla}_\perp\phi).
\label{enden}
\ee

The force density is given by
\be
f_\lambda=-\frac1{\sqrt{-g}}\partial_\nu(\sqrt{-g}T^\nu{}_\lambda)
+\frac12T^{\mu\nu}\partial_\lambda g_{\mu\nu},
\ee
or in Rindler coordinates
\be
f_\xi=-\frac1\xi\partial_\xi(\xi T^{\xi\xi})-\xi T^{00}.\label{fxi}
\ee
When we integrate over all space to get the force, the first term is
a surface term which does not contribute:\footnote{Note that in previous
works, such as Refs.~\cite{Milton:2004vy,Milton:2004ya}, the surface
term was included, because the integration was carried out only over
the interior and exterior regions.  Here we integrate over the surface
as well, so the additional so-called surface energy is automatically 
included.  Note, however, if Eq.~(\ref{fxi}) is integrated over a
small interval enclosing the $\delta$-function potential,
$$
\int_{\xi_1-\epsilon}^{\xi_1+\epsilon} d\xi\,\xi f_\xi=-\xi_1\Delta T^{\xi\xi},
$$
where $\Delta T^{\xi\xi}$ is the discontinuity in the normal-normal
component of the stress density.  Dividing this expression by $\xi_1$ gives
the usual expression for the force on the plate.}
\be
\mathcal{F}=\int d\xi\, \xi f_\xi=-\int\frac{d\xi}{\xi^2}T_{00}.
\label{ep}
\ee
This could be termed the Rindler coordinate force per area, defined as the
change in momentum per unit Rindler coordinate time $\tau$ per unit
cross-sectional area.  If we 
multiply $\mathcal{F}$  by the gravitational acceleration $g$ we obtain
the gravitational force per area on the Casimir energy.
This result (\ref{ep}) seems entirely consistent with the equivalence
principle, since $\xi^{-2}T_{00}$ is the energy density.
Using the expression (\ref{enden}) for the energy density, 
taking the vacuum expectation value, and rescaling
$\zeta=\hat\zeta\xi$,
we see that the gravitational force per cross sectional area is merely
\be
\mathcal{F}=\int d\xi \,\xi\int\frac{d\hat\zeta \, d^2\mathbf{k}}
{(2\pi)^3}\hat\zeta^2g(\xi,\xi).\label{gravf-gf}
\ee

This result for the energy contained in the force equation (\ref{gravf-gf})
is an immediate consequence of the general formula
for the Casimir energy \cite{Milton:2001yy}
\be
E_c=-\frac1{2i}\int(d\mathbf{r})\int\frac{d\omega}{2\pi}
2\omega^2\mathcal{G}(\mathbf{r,r}),
\ee
in terms of the frequency transform of the Green's function,
\be
G(x,x')=\int_{-\infty}^\infty\frac{d\omega}{2\pi}e^{-i\omega(t-t')}\mathcal{
G}(\mathbf{r,r'}).
\ee

Alternatively, we can start from the following formula for the force
density for a single semitransparent plate, following directly from
the equations of motion (\ref{eom}),
\be
f_\xi=\frac12\phi^2\partial_\xi \lambda\delta(\xi-\xi_1).\label{fd}
\ee
The vacuum expectation value of this yields the force 
in terms of the Green's function,
\be
\mathcal{F}=-\lambda\frac12
\int\frac{d\zeta\,d^2\mathbf{k}}
{(2\pi)^3}\partial_{\xi_1}[\xi_1 g(\xi_1,\xi_1)].
\ee

\subsection{Gravitational force on a single plate}

For example, the force on a single plate at $\xi_1$ is given by
\be
\mathcal{F}=-\partial_{\xi_1} \frac12\int\frac{d\zeta\,d^2\mathbf{k}}
{(2\pi)^3}\ln[1+\lambda \xi_1 I_\zeta(k\xi_1)K_\zeta(k\xi_1)],
\ee
Expanding this about some arbitrary point $\xi_0$, with $\zeta=\hat\zeta\xi_0$,
using the uniform asymptotic expansion (\ref{uae}), we get 
\be
\xi_1 I_\zeta(k\xi_1)K_\zeta(k\xi_1)\sim\frac{\xi_1}{2\zeta}\frac1
{\sqrt{1+(k\xi_1/\zeta)^2}}\approx\frac{\xi_1}{2\kappa\xi_0}\left(1-\frac{k^2}
{\kappa^2}\frac{\xi_1-\xi_0}{\xi_0}\right).
\ee
From this, if we introduce polar coordinates for the $\mathbf{k}$-$\hat\zeta$ 
integration ($\kappa^2=k^2+\hat\zeta^2$), the
coordinate force is
\bea
\mathcal{F}&=&-\frac12\partial_{\xi_1}\frac{\xi_0}{2\pi^2}\int_0^\infty
d\kappa\,\kappa^2\frac{\lambda}{2\kappa+\lambda}
\left(1+\frac{\xi_1-\xi_0}{\xi_0}
\right)\left(1-\frac{\langle k^2\rangle}{\kappa^2}\frac{\xi_1-\xi_0}{\xi_0}
\right)\nn\\
&=&-\frac\lambda{4\pi^2}\partial_{\xi_1}(\xi_1-\xi_0)\int_0^\infty
\frac{d\kappa}{2\kappa+\lambda}\langle\hat\zeta^2\rangle
\nn\\
&=&-\frac{1}{96\pi^2a^3}\int_0^\infty\frac{dy\,y^2}{1+y/\lambda a},
\label{singleplate}
\eea
where for example
\be
\langle\hat\zeta^2\rangle=\frac12\int_{-1}^1 d\cos\theta\, \cos^2\theta\, 
\kappa^2=\frac13\kappa^2.
\ee
The divergent expression (\ref{singleplate}) 
 is just the negative of the quantum vacuum energy of a single plate.


\subsection{Parallel plates falling in a constant gravitational field}

In general, we have two alternative forms for the 
gravitational force on the two-plate system:
\be
\mathcal{F}=-(\partial_{\xi_1}+\partial_{\xi_2})\frac12\int\frac{d\zeta\,
d^2\mathbf{k}}{(2\pi)^3}\ln\Delta,\ee
$\Delta$ given in Eq.~(\ref{Delta}),
which is equivalent to (\ref{gravf-gf}).  (In the latter, however,
bulk energy, present if no plates are present, must be omitted.)
From either of the above two methods, we find the coordinate force
is given by
\be
\mathcal{F}=-\frac1{4\pi^2}\int_0^\infty d\kappa\,\kappa^2 \ln\Delta_0,
\label{coordf}
\ee
where $\Delta_0=e^{-2\kappa a}\tilde\Delta$, $\tilde\Delta$ given in
Eq.~(\ref{tDelta}).
The integral may be easily shown to be
\begin{subequations}
\label{casen}
\bea
\mathcal{F}&=&\frac1{96\pi^2 a^3}\int_0^\infty dy\,y^3\frac{1+
\frac1{y+\lambda_1a}
+\frac1{y+\lambda_2a}}{\left(\frac{y}{\lambda_1a}+1\right)
\left(\frac{y}{\lambda_2a}+1\right)e^y-1}
-\frac1{96\pi^2 a^3}\int_0^\infty dy\,y^2\left[\frac1{\frac{y}{\lambda_1a}
+1}+\frac1{\frac{y}{\lambda_2a}+1}\right]\label{negce}\\
&=&-(\mathcal{E}_c+\mathcal{E}_{d1}+\mathcal{E}_{d2}),\label{fisen}\eea
\end{subequations}
which is just the negative of the Casimir energy of the two semitransparent
plates including the divergent pieces \cite{Milton:2004vy,Milton:2004ya}.
Note that $\mathcal{E}_{di}$, $i=1,2$, are simply the divergent energies
(\ref{singleplate}) associated with a single plate.
 
\subsection{Renormalization}

The divergent terms in Eq.~(\ref{casen}) 
simply renormalize the masses (per unit area) of each plate:
\bea
E_{\rm total}&=&m_1+m_2+\mathcal{E}_{d1}+\mathcal{E}_{d2}+\mathcal{E}_c\nn\\
&=&M_1+M_2+\mathcal{E}_c,
\eea
where $m_i$ is the bare mass of each plate, and the renormalized mass
is $M_i=m_i+\mathcal{E}_{di}$.
Thus the gravitational force on the entire apparatus obeys the
equivalence principle
\be
g\mathcal{F}=-g(M_1+M_2+\mathcal{E}_c).
\ee
The minus sign reflects the downward acceleration of gravity on the
surface of the earth.  Note here that the Casimir interaction energy
$\mathcal{E}_c$ is negative, so it reduces the gravitational attraction
of the system.

\section{Conclusions}

We have found, in conformation with the result given in 
Ref.~\cite{Fulling:2007xa}, an extremely simple answer to the question of
how Casimir energy accelerates in a weak gravitational field: 
Just like any other form of energy, the gravitational force $F$ divided
by the area of the plates is
\be
\frac{F}A=-g\mathcal{E}_c.
\ee
This is the result expected by the equivalence principle, but
is in contradiction to some earlier disparate claims in the 
literature \cite{karim,calloni,caldwell,Sorge:2005ed,bimonte}.
This result exactly agrees with that found by Saharian 
et al.~\cite{Saharian:2003fd}
for Dirichlet, Neumann, and perfectly conducting plates for the finite
Casimir interaction energy.  The acceleration of Dirichlet plates follows from
our result when the strong coupling limit $\lambda\to\infty$ is taken.
What makes our conclusion particularly interesting is that it refers not
only to the finite part of the Casimir interaction energy between 
semitransparent plates, but to the divergent parts as well, which are
seen to simply renormalize the gravitational mass of each plate, as they
would the inertial mass.  The reader may object that by equating gravitational
force with uniform acceleration we have built in the equivalence principle,
and so does any procedure based on Einstein's equations; but the real
nontriviality here is that quantum fluctuations obey the same universal
law.

\begin{acknowledgments}
We are grateful to S. Fulling and A. Romeo for extensive collaborative advice. 
We would like to thank R. Kantowski for formal and informal
discussions.
KVS would like to thank David S. Hartnett for helpful conversations.
This work was supported in part by grants from the US National Science
Foundation (PHY-0554926) and the US Department of Energy (DE-FG02-04ER41305).
\end{acknowledgments}

\appendix
\section{Casimir and gravitational forces}
In this paper we have concentrated on the gravitational acceleration
on a rigid Casimir apparatus, and not on the Casimir forces between
the plates.  It may be useful to recognize that both of these forces
are contained within the formalism given.  So let us consider the
force on one plate of a two plate apparatus, as described by the Green's
function given in Sec.~\ref{sec2c}.  According to Eq.~(\ref{fd}), the
force on plate 2 is given by
\be
\mathcal{F}_2=-\partial_{\xi_2}\frac12\int\frac{d\zeta\,
d^2\mathbf{k}}{(2\pi)^3}\ln\Delta,
\ee
where $\Delta$ is given by Eq.~(\ref{Delta}).  We expand the latter in
inverse powers of an arbitrary reference point $\xi_0\gg1$:
\be
\Delta(\xi_1,\xi_2)=\Delta_0(\xi_2-\xi_1)+\frac1{\xi_0}\Delta_1(\xi_1,\xi_2)
+O(\xi_0^{-2}).
\ee
Then, in view of Eqs.~(\ref{fisen}) and (\ref{coordf}), 
the coordinate force on plate 2 is ($\hat\zeta=\zeta/\xi_0$)
\be
\mathcal{F}_2=-\xi_0\frac\partial{\partial a}\mathcal{E}_c+\frac12
\partial_{\xi_2}\int\frac{d\hat\zeta\,d^2\mathbf{k}}{(2\pi)^3}\frac{\Delta_1}
{\Delta_0}(\xi_1,\xi_2)+O\left(\frac1{\xi_0}\right).
\ee
The first term here, when divided by $\xi_0$ to give the physical force,
coincides with the usual expression for the Casimir force 
on one semitransparent
plate due to a second plate a distance $a$ away.  The second term, when
multiplied by $g=1/\xi_0$, is the gravitational force on the second plate.
When the corresponding forces on plate 1 are added to these, the
Casimir forces cancel, while the sum of the gravitational forces is seen
to be exactly that given in Eq.~(\ref{negce}).

\end{document}